\documentclass[aps,prb,showpacs]{revtex4}
\usepackage{bezier}
\usepackage{graphicx}
\usepackage{bm}
\usepackage{amssymb}
\usepackage{amsmath}

\begin{document}
\title{Operator method for solution of
the Schr\"{o}dinger equation
with the rational potential}
\author{Petr A. Khomyakov}

\address{Department of Theoretical Physics,
Belarussian State University,
F.\ Skarina av.\ 4, Minsk 220080, Republic of Belarus}
\date{\today}

\begin{abstract}
The eigenvalue problem for one-dimensional Schr\"{o}dinger equation
with the rational potential is numerically solved by the operator
method. We show that the operator method, applied for solving the
Schr\"{o}dinger equation with the nonpolynomial structure of the
Hamiltonian, becomes more efficient if a nonunitary transformation
of the Hamiltonian is used. We demonstrate on numerous examples that
this method can handle both perturbative and nonperturbative regimes
with very high accuracy and moderate computational cost.
\end{abstract}
\maketitle

\section{Introduction}
In this paper we generalize the operator method for solving the
Schr\"{o}dinger equation (SE) with a nonpolynomial structure of the
Hamiltonian. As an example of such a problem we consider
one-dimensional Hamiltonian with the rational potential
\begin{equation}\label{f2}
{{V}}_{L} (x) = \frac{{x}^{2}}{2} +
\frac{\lambda {x}^{2L}}{1 + g {x}^{2}}.
\end{equation}
The problem chosen is of interest in different areas of physics,
i.e. laser physics \cite{l15} and non-linear quantum field theory
\cite{l16}. It can be also exploited for verifying new
nonperturbative methods, which allow us to find approximate solution
of the Schr\"{o}dinger equation
 in the whole range of Hamiltonian parameters. Such methods can be
 used
for treating many physical problems of current interest. The test
problem chosen is very suitable for these purposes because the SE
\begin{equation}\label{f1}
\left[ \frac{ {p}^{2}}{2} + {V}_{L}
({x}) - E \right] \vert \psi \rangle = 0,
\end{equation}
can be solved exactly if some algebraic relations between parameters
$g$ and $\lambda$ are assumed \cite{l11,l17,l12}. The exact
solutions allow to find out the accuracy of approximate methods.

The Schr\"{o}dinger equation with the potentials $V_{1} (x)$ and
$V_{2} (x)$ has been treated by different approaches based on
finite-difference schemes, variational and asymptotic
methods\cite{l10,l3,l2,l7,l8,l9,l1,l4}. To the best of our knowledge
there is only one work \cite{l4}, where the case $L>2$ was studied
in the wide range of the parameter $\alpha=g/\lambda$ by very
efficient method of finite-differences.
 The case $L=+3$ has been studied for $g\ll 1,\lambda\ll 1$ by
using the Pad\'e-approximant and hypervirial methods\cite{l20}.
Three-dimensional generalization of Eq.(\ref{f1}) can be found in
Ref.~\onlinecite{l19}.

In the present paper the SE Eq.(\ref{f1}) is solved by the operator
method (OM), developed by Komarov and Feranchuk\cite{l5}. The OM has
been successfully applied to different
problems\cite{l18,thu,l21,l14}, but the most problems considered had
a polynomial structure of the Hamiltonian. The polynomial structure
allows one to calculate matrix elements of the SE equation
analytically in the Fock basis\cite{l1}. But this simple analytical
scheme can not be directly applied to the nonpolynomial equation
Eq.(\ref{f1}). We can still perform the OM calculations\cite{l18},
but some extra numerical scheme must be exploited to calculate
matrix elements. This problem can be resolved by the use of the
nonunitary transformation $C$
\begin{equation}\label{f6}
\vert\psi\rangle = C\; \vert\varphi\rangle, \qquad
C = (1 + g {x}^{2}),
\end{equation}
which leads to the following expression for Eq.(\ref{f1})
\begin{equation}\label{f7}
\left[ \frac{{p}^{2}}{2} + \frac{{x}^{2}}{2} +
g \left( \frac{{p}^{2} {x}^{2}}{2} +
\frac{{x}^{4}}{2} \right) + \lambda{x}^{2L}
- (1 + g {x}^{2}) E \right] \vert \varphi \rangle = 0.
\end{equation}
So we transformed the nonpolynomial operator equation Eq.(\ref{f1})
to the polynomial one Eq.(\ref{f7}), where the operator at the left
side is nonhermitian with respect to the scalar product
\begin{equation}
\langle\phi\vert\psi\rangle =
\int {\rm d}{{x}} \, \phi^{*} ({x}) \psi ({x}).
\end{equation}
The idea of using nonunitary transformations was recently suggested
in Ref.~\onlinecite{ander}. It was also developed in
Ref.~\onlinecite{lee} as nonhermitian technique of canonical
transformations. The basic idea is that the commutation relation $[
x , p ] = i$ can be preserved by any unitary transformation, but the
same is also true if we make use of the similarity transformation $X
= C\, x\, C^{-1}$, $P = C\, p\, C^{-1}$. Below we show that the
polynomial operator equation Eq.(\ref{f7}) is very easy to handle
within the framework of the operator method in order to find both
the exact numerical and approximate analytical solutions. The OM
allows us to treat the problem in the wide range of Hamiltonian
parameters $g$ and $\lambda$, where the SE Eq.(\ref{f1}) has a
discrete energy spectrum, i.e. when $-\infty < L\leq +1 $,
$\vert\lambda\vert < +\infty$, $g\geq 0$; $L> +2$, $\lambda\geq 0$,
$g\geq 0$; $L=+2$, $\lambda > -g/2$, $g\geq 0$. Below we present
results of our nonperturbative calcutations for four cases $1\leq L
\leq 4$.

\section{Iterative scheme of the operator method}
 In this section we apply the operator method for solving the equation Eq.(\ref{f7})
 using the scheme developed in Ref.~\onlinecite{l5}. If
we introduce the annihilation operator $a(\omega)$, the creation
operator $a^{+}(\omega)$ and the excitation number operator
${n}(\omega) = a^{+}(\omega)a(\omega)$
\begin{equation}\label{f4}
\displaystyle{ a^{+}(\omega) =
\frac{1}{\sqrt{2\omega}} (\omega{x} - i {p})},
\qquad
\displaystyle{a(\omega) = \frac{1}{\sqrt{2\omega}}
(\omega{x} + i {p})},
\end{equation}
which satisfy the commutation relations
\begin{equation}\label{f5}
[ a(\omega), a^{+}(\omega) ] = 1, \qquad
[ a(\omega), {n}(\omega) ] = a(\omega), \qquad
[ a^{+}(\omega), {n}(\omega) ] = - a^{+} (\omega),
\end{equation}
one have for Eq.(\ref{f7})
\begin{eqnarray}\label{f8}
({\cal L}^{\prime} &-& E\,
{\cal L}^{\prime\prime}) \vert\varphi\rangle = 0,
\\
{\cal L}^{\prime} &=&
\left(\frac{1}{4\omega} + \frac{\omega}{4}\right)
(2 {n} + 1) +
\frac{3g}{8 \omega^{2}}(2 {n}^{2} + 2{n} + 1) +
\frac{g}{8}(2{n}^{2} + 2{n} - 1)
+ \frac{g}{8}(\frac{1}{\omega^{2}} - 1) \nonumber
\\
&\times& (a^{+4} + a^{4}) +
\left( \frac{1}{4\omega} - \frac{\omega}{4} +
\frac{3g}{4\omega^{2}} + \frac{g}{2\omega^{2}}
{n}\right)(a^{+2}+a^{2})
- \frac{g}{\omega^{2}}a^{+2} \nonumber
\\
&+& \frac{g}{2} (a^{+2} - a^{2}) +
\lambda {x}^{2L},
\\
{\cal L}^{\prime\prime} &=& 1 +
\frac{g}{2\omega}(a^{+2} + a^{2} + 2{n} + 1),
\end{eqnarray}
where ${x}^{2L}$ is written below for the four particular cases
$1\leq L \leq 4$
\begin{eqnarray}\label{f9}
{x}^{2} &=&
\frac{1}{2\omega}(2 {n} + 1) +
\frac{1}{2\omega}(a^{+2} + a^{2}), \\
{x}^{4} &=& \frac{3}{4\omega^{2}}
(1 + 2 {n} + 2 {n}^{2})
+\frac{1}{4\omega^{2}}(a^{+4} + a^{4} +
(4 {n} - 2) a^{+2} + (4{n} + 6) a^{2}),
\\
{x}^{6} &=& \frac{5}{8\omega^{3}}(3 + 8{n} +
6 {n}^{2} + 4 {n}^{3}) +
\frac{1}{8\omega^{3}}
(a^{+6} + a^{6} + (6{n} - 9) a^{+4} + (6{n} + 15) a^{4}
\nonumber
\\
&& +
15 (1 - {n} + {n}^{2}) a^{+2} +
15 (3 + 3{n} + {n}^{2})a^{2}),
\\
{x}^{8} &=& \frac{1}{16\omega^{4}}
(105 + 280{n} + 350{n}^{2} +
140 {n}^{3} + 70 {n}^{4}) \nonumber
\\
&& + \frac{1}{16 \omega^{4}}(a^{+8} + a^{8} +
(8 {n} - 20) a^{+6} + (8{n} + 28) a^{6} \nonumber
\\
&& + (98 - 84 {n} + 28 {n}^{2}) a^{+4} +
(210 + 140 {n} + 28 {n}^{2}) a^{4} \nonumber
\\
&& + (-84 + 196 {n} - 84 {n}^{2} +
56 {n}^{3}) a^{+2} +
(420 + 532 {n} + 252{n}^{2} + 56 {n}^{3}) a^{2}).
\end{eqnarray}
Any degree of ${x}^{2}$ can be derived by the use of a recursive
relation between ${x}^{2L}$ and ${x}^{2L-2}$. The normalized
eigenvectors $\vert n,\omega\rangle$ of the excitation number
operator $n$ are
\begin{equation}\label{f3}
\vert n,
\omega\rangle = \frac{ \left({a^{+}(\omega)}\right)^{n} }
{\sqrt{n!}} \vert 0, \omega \rangle, \qquad
a( \omega ) \vert 0, \omega \rangle = 0.
\end{equation}
In contrast to the equation Eq.(\ref{f1}) we can calculate the
matrix elements of the equation Eq.(\ref{f8}) exactly in the Fock
basis Eq.(\ref{f3}). The eigenvector $\vert\varphi\rangle$ can be
represented in the Fock basis Eq.(\ref{f3}) as
\begin{equation}\label{f10}
\vert \varphi \rangle =
\sum_{p=0}^{\infty} {C}_{p} \vert p,\omega \rangle,
\end{equation}
with the coefficient $C_{p}$ to be defined by the equation
Eq.(\ref{f8}). The arguments of $a^{+}(\omega)$, $a(\omega)$,
$n(\omega)$ and $\vert n,\omega\rangle$ will be omitted in order to
simplify all expressions. Substitute Eq.(\ref{f10}) in Eq.(\ref{f8})
and find its projection on bra-vector $\langle k,\omega\vert$
\begin{equation}\label{f11}
{C}_{k} \langle k \vert
{\cal L}^{\prime} \vert k \rangle -
{C}_{k} \langle k \vert {\cal L}^{\prime\prime}\vert k\rangle {E}
+\sum_{p\neq k} {C}_{p}
\langle k\vert{\cal L}^{\prime}\vert p\rangle -
{E} \sum_{p\neq k} {C}_{p}
\langle k \vert {\cal L}^{\prime\prime} \vert p\rangle = 0.
\end{equation}
The system of the linear equations Eq.(\ref{f11}) can be solved by
the iterative method\cite{l14}
\begin{eqnarray}\label{f12}
&& {C}^{(s)}_{k,k\neq n}=-\frac{\sum_{p=k-m,p\neq k}^{k+m}{C}^{(s-1)}_p
\langle k\vert
{{\cal L}}^{\prime}\vert p\rangle-{E}^{(s-1)}_{{n}}
\sum_{p=k-m,p\neq k}^{k+m}{C}^{(s-1)}_p\langle k\vert
{{\cal L}}^{\prime\prime}\vert p\rangle}
{\langle k\vert
{{\cal L}}^{\prime}\vert k\rangle-{E}^{(s-1)}_{{n}}
\langle k\vert {{\cal L}}^{\prime\prime}
\vert k\rangle}, \nonumber \\
&&{E}^{(s)}_{{n}}=\frac{\langle {n}\vert
{{\cal L}}^{\prime}\vert {n}\rangle+
\sum_{p={n}-m,p\neq n}^{{n}+m} {C}^{(s-1)}_p\langle {n}\vert
{{\cal L}}^{\prime}\vert p\rangle}
{\langle {n}\vert
{{\cal L}}^{\prime\prime}\vert {n}\rangle+
\sum_{p={n}-m,p\neq n}^{{n}+m}
{C}^{(s-1)}_p
\langle {n}\vert {{\cal  L}}^{\prime\prime}\vert  p\rangle},\qquad
{C}^{(0)}_{{k}}=\delta_{n,k}
\end{eqnarray}
where $C^{(s)}_{k}$ is the coefficient of the eigenvector expansion
and ${E}^{(s)}_{n}$ is the energy of $n$-level calculated in $s$-th
iteration; $m=2L$; $s=0,1, \ldots ,{s}_{{\rm max}}$. The results of
numerical calculations of the energy eigenvalues for the ground
($n=0$) and first excited ($n=1$) states in the wide range of values
of $g$, $\lambda$ are summarized in Tables 1-4. For convenience of
the comparison with the results in Refs.~\onlinecite{l1,l20} we
present our results in the form $2\lambda$, $2E$. As is clear from
Table 1 and Table 2, where some exact eigenvalues are given, the
accuracy of the operator method is very high $\vert E^{(s_{\rm
max})}_{n} - E_{n}\vert/E_{n} < 10^{-15}$. But the case $g\gg 1$,
$\lambda\gg 1$ needs the increase of the iteration number $s_{\rm
max}$. So it is important to make use of the parameter $\omega$,
which allows us to speed up the convergence of the iterative scheme.
Since the exact eigenvalues of the Hamiltonian do not depend on the
choice of $\omega$, the following condition has to be satisfied for
exact energy eigenvalues\cite{l5}
\begin{equation}\label{12}
\displaystyle{\frac{\partial {E}_{n}}{\partial\omega}=0.}
\end{equation}
Because $E_{n}$ is equal to ${E}^{(\infty)}_{n}(\omega)$, a good
accuracy can be achieved at the extremum point $\omega_{extr}$ of
the function ${E}^{(s_{\rm max})}_{n}(\omega)$ (see Fig. 1). Notice
that this statement is only valid in the case of large iteration
number. As soon as the iteration number is relatively small
${E}^{(s_{\rm max})}_{n}(\omega)$ is described by an oscillating
function crossing $E=E_{n}({\rm exact})$ for some values of $\omega$
(see Fig. 1 and Ref.~\onlinecite{l21}). In the case of large
iteration number the magnitude of oscillations is very small and
goes to zero quickly (see Fig. 1).

We emphasize that even for small number of iterations the points of
extremum are near the exact energy eigenvalue and belong to the
range of stable convergence of the iterative scheme. They can be
chosen as starting values for $\omega$. As is clear from Table 2 the
values of ${s}_{\rm max}$ and $\omega$ must be increased for excited
states. Large values of $g$ and $\lambda$ require the increase of
${\omega}$ as well. Such behavior is typical for all potentials $
V_{L}(x) $. Different criteria for choosing $\omega$ can be also
found in Refs.~\onlinecite{l5,l18,l21}.

We'd like to notice that the normalized eigenfunctions
${\Psi}_{n}(x)$ of the original SE Eq.(\ref{f1}) can be easily
calculated
\begin{eqnarray}
&& \Psi_{n}(x) =
\frac{{\psi}_{n}(x)}{\sqrt{\langle \psi_{n} \vert \psi_{n}\rangle}}
\approx
\frac{1}{N}(1 + g x^{2})\sum_{k = 0}^{{s}_{\rm max}}
{C}^{(s_{\rm max})}_{k}
\langle x\vert k,{\omega}\rangle
\\
&& N^{2} = \langle \psi_{n} \vert \psi_{n} \rangle
\approx  \int^{-\infty}_{+\infty} {\rm d}x
(1 + g x^{2})^{2} \left[ \sum_{k = 0}^{{s}_{\rm max}}
{C}^{(s_{\rm max})}_{k}
\langle x \vert k, {\omega} \rangle \right]^{2}.
\end{eqnarray}
The polynomial structure of the operator equation Eq.(\ref{f7}) also
allows us to obtain some analytical expressions for the energy
eigenvalues with the explicit dependence on all parameters. Such
expressions can be found by making a limited number of iterations
with any system of computer algebra like {\it Maple} or {\it
Mathematica}.

\section{Conclusion}
The Schr\"{o}dinger equation with the rational potential was solved
by the operator method. The energy of the ground state and the first
excited state were calculated with very high accuracy without any
limitations on the parameters of the Hamiltonian. The numerical
solutions are in excellent agreement with the exact solutions known
for some values of $g$ and $\lambda$. The criteria for choosing the
convergence parameter $\omega$ was analyzed. So we have shown that
the spectrum of the original nonpolynomial Schr\"{o}dinger equation
Eq.(\ref{f1}) is identical to the one of the polynomial operator
equation Eq.(\ref{f7}).
 The results obtained demonstrate that the operator method
is very efficiently for solving the problems with the nonpolynomial
structure of the Hamiltonian if the nonunitary transformation is
used. The application of the operator method to the
three-dimensional nonpolynomial Hamiltonian is straightforward.

\section*{Acknowledgments}
The author is grateful to L.~I.~Komarov for useful discussions.

\newpage
\begin{table}
\caption{Energy eigenvalues for $V_{1}(x)$.}
\begin{center}
\begin{tabular}{| llll |}
\hline
$2\lambda/g$ & 100/10 & 10/10 & 10/100\\
\hline
$2E_{0}(OM)$ & 5.79394230019270 & 1.58002232739150
& 1.08406333549441 \\
$2E_{1}(OM)$ & 11.5721967757092 & 3.87903683088257
& 3.09831699508894\\
\hline
$2\lambda/g$ & $-2.494002/0.499$ & $-1.585842/0.339$
& $-0.0402/0.01$\\
\hline
$2E_0$(OM)&0.002000000000000&0.322000000000000
&0.98000000000000\\
$2E_0$(exact)&0.002000000000000&0.322000000000000
&0.98000000000000\\
\hline
\end{tabular}
\end{center}
\end{table}
\begin{table}
\caption{Energy eigenvalues for $V_{2}(x)$.}
\begin{center}
\begin{tabular}{| llll |}
\hline
$2\lambda/g$ & 0.1/1 & 0.1/0.1 & 1/0.1 \\
\hline
$2E_0$(OM) & 1.02514716380525 & 1.05529770725788
& 1.36059173241772\\
$ \omega / s_{\rm max} $& 3.0/69 & 1.74/20 & 2.90/22 \\
$2E_1$(OM) & 3.09577659167734 & 3.24865072344421
& 4.49215757165382\\
$ \omega / s_{\rm max} $ & 2.9/70 & 1.56/20 & 2.82/22
\\\hline
$2\lambda/g$ & 100/10 & 10/10 & 10/100\\
\hline
$2E_{0}$ (OM) & 2.90257776978742 & 1.35977466215783
& 1.04797115328336  \\
$ \omega / s_{\rm max} $ & 16.6/115 & 11.0/200 &
27.3/1400 \\
$2E_{1}$ (OM) & 9.21956081849447 & 4.15798607191020 & 3.14544411596743 \\
$ \omega / s_{\rm max} $ & 16.3/125 & 10.4/210 & 25.7/1700 \\
\hline
$\lambda/g$ & 0.001/0.02 & 0.005/0.1 & 0.01/0.01\\
\hline
$E_0$(OM) & 0.500712692243485 & 0.503010339434091
& 0.507093239851910\\
$E_0$(\cite{l20})& 0.5007126922434854& 0.503010
& 0.507093239851\\
$E_1$(OM)& 1.503492319497928 & 1.51392272439619
& 1.53457040870676\\
$E_1$(\cite{l20})& 1.5034923194979289 & 1.51392
& 1.534570408\\
\hline
$\alpha/g$ & $0.5215/0.2607543752089$ &  &\\
\hline
$2E_0$(OM)&1.18657366612601 & & \\
$2E_0$(exact)&1.18657366612601 &  &\\
\hline
\end{tabular}
\end{center}
\end{table}
\begin{table}
\caption{Energy eigenvalues for $V_{3}(x)$.}
\begin{center}
\begin{tabular}{| llll |}
\hline
$2\lambda/g$ & 100/10 & 10/10 & 10/100\\
\hline
$2E_{0}$ (OM) & 2.32812938769652 & 1.36685074612832
& 1.06484795071250 \\
$2E_{1}$ (OM) & 8.25858912303721 & 4.57644754780293
& 3.30561773097224 \\
\hline
$\lambda/g$&  $10^{-4}/10^{-3}$&$10^{-4}/0.01$&
$5\times 10^{-4}/2.5\times 10^{-3}$\\
\hline
$E_0$(OM)& 0.500186311611817 & 0.500180733817835
& 0.500916657050025\\
$E_0$(\cite{l20}) & 0.5001863116118168 & 0.5001807336178352
& 0.5009166570\\
\hline
$2\lambda/g$&10/0 & 1000/0 & 10000/0 \\
\hline
$2E_0$(OM)& 2.20572326959563 & 6.49235013232967
& 11.4787980422645 \\
$2E_0$(\cite{l4})& 2.205723269595632 & 6.492350132329672 &
11.47879804226454 \\
\hline
\end{tabular}
\end{center}
\end{table}
\begin{table}
\caption{ Energy eigenvalues for $V_{4}(x)$.}
\begin{center}
\begin{tabular}{| llll |}
\hline
$2\lambda/g$ & 100/10 & 10/10 & 10/100\\
\hline
$2E_{0}$ (OM) & 2.14407075597610 & 1.41752810562730
& 1.10864509823789 \\
$2E_{1}$ (OM) & 3.59432196739250 & 4.96949962368201
& 3.59432196739250  \\
\hline
$2\lambda/g$& 10/0 & 1000/0 & 10000/0 \\
\hline
$2E_0$(OM)& 2.11454462194213 & 4.94948744003274
& 7.77827221431110 \\
$2E_0$(\cite{l4})& 2.114544621942129 & 4.949487440032743
& 7.778272214311099 \\
\hline
\end{tabular}
\end{center}
\end{table}

\newpage
\begin{figure}[]
\thispagestyle{empty}
\begin{center}
\begin{picture}(315,315)(0,0)
\put(0,0){\vector(4,0){315}}
\put(0,0){\vector(0,4){315}}
\put(0,-2){\line(0,6){4}}
\put(-2,-10){\scriptsize 0}
\put(60,-2){\line(0,6){4}}
\put(58,-10){\scriptsize 1}
\put(120,-2){\line(0,6){4}}
\put(118,-10){\scriptsize 2}
\put(180,-2){\line(0,6){4}}
\put(178,-10){\scriptsize 3}
\put(240,-2){\line(0,6){4}}
\put(238,-10){\scriptsize 4}
\put(300,-2){\line(0,6){4}}
\put(298,-10){\scriptsize 5}
\put(-2,0.00){\line(6,0){4}}
\put(-14,-2.00){\scriptsize 1.0}
\put(-2,60.00){\line(6,0){4}}
\put(-14,58.00){\scriptsize 1.2}
\put(-2,120.00){\line(6,0){4}}
\put(-14,118.00){\scriptsize 1.4}
\put(-2,180.00){\line(6,0){4}}
\put(-14,178.00){\scriptsize 1.6}
\put(-2,240.00){\line(6,0){4}}
\put(-14,238.00){\scriptsize 1.8}
\put(-2,300.00){\line(6,0){4}}
\put(-14,298.00){\scriptsize 2.0}
\put(320,-2){$\omega$}
\put(-2,320){$2E^{(s)}_{0}$}
\put(260,300){$s=0$}
\put(290,180){$s=3$}
\put(290,65){$s=\infty$}
\bezier{50}(0.00,55.97)(6.00,55.97)(12.00,55.97)
\bezier{50}(12.00,55.97)(18.00,55.97)(24.00,55.97)
\bezier{50}(24.00,55.97)(30.00,55.97)(36.00,55.97)
\bezier{50}(36.00,55.97)(42.00,55.97)(48.00,55.97)
\bezier{50}(48.00,55.97)(54.00,55.97)(60.00,55.97)
\bezier{50}(60.00,55.97)(66.00,55.97)(72.00,55.97)
\bezier{50}(72.00,55.97)(78.00,55.97)(84.00,55.97)
\bezier{50}(84.00,55.97)(90.00,55.97)(96.00,55.97)
\bezier{50}(96.00,55.97)(102.00,55.97)(108.00,55.97)
\bezier{50}(108.00,55.97)(114.00,55.97)(120.00,55.97)
\bezier{50}(120.00,55.97)(126.00,55.97)(132.00,55.97)
\bezier{50}(132.00,55.97)(138.00,55.97)(144.00,55.97)
\bezier{50}(144.00,55.97)(150.00,55.97)(156.00,55.97)
\bezier{50}(156.00,55.97)(162.00,55.97)(168.00,55.97)
\bezier{50}(168.00,55.97)(174.00,55.97)(180.00,55.97)
\bezier{50}(180.00,55.97)(186.00,55.97)(192.00,55.97)
\bezier{50}(192.00,55.97)(198.00,55.97)(204.00,55.97)
\bezier{50}(204.00,55.97)(210.00,55.97)(216.00,55.97)
\bezier{50}(216.00,55.97)(222.00,55.97)(228.00,55.97)
\bezier{50}(228.00,55.97)(234.00,55.97)(240.00,55.97)
\bezier{50}(240.00,55.97)(246.00,55.97)(252.00,55.97)
\bezier{50}(252.00,55.97)(258.00,55.97)(264.00,55.97)
\bezier{50}(264.00,55.97)(270.00,55.97)(276.00,55.97)
\bezier{50}(276.00,55.97)(282.00,55.97)(288.00,55.97)
\bezier{50}(288.00,55.97)(294.00,55.97)(300.00,55.97)
\bezier{50}(48.00,177.57)(54.00,123.45)(60.00,99.52)
\bezier{50}(60.00,99.52)(66.00,74.73)(72.00,64.69)
\bezier{50}(72.00,64.69)(78.00,54.34)(84.00,52.13)
\bezier{50}(84.00,52.13)(90.00,49.78)(96.00,52.35)
\bezier{50}(96.00,52.35)(102.00,54.84)(108.00,60.50)
\bezier{50}(108.00,60.50)(114.00,66.11)(120.00,73.86)
\bezier{50}(120.00,73.86)(126.00,81.60)(132.00,90.83)
\bezier{50}(132.00,90.83)(138.00,100.05)(144.00,110.37)
\bezier{50}(144.00,110.37)(150.00,120.68)(156.00,131.80)
\bezier{50}(156.00,131.80)(162.00,142.93)(168.00,154.68)
\bezier{50}(168.00,154.68)(174.00,166.42)(180.00,178.66)
\bezier{50}(180.00,178.66)(186.00,190.90)(192.00,203.52)
\bezier{50}(192.00,203.52)(198.00,216.14)(204.00,229.08)
\bezier{50}(204.00,229.08)(210.00,242.02)(216.00,255.21)
\bezier{50}(216.00,255.21)(222.00,268.40)(228.00,281.81)
\bezier{50}(228.00,281.81)(234.00,295.21)(240.00,308.80)
\bezier{50}(39.00,99.90)(45.00,70.45)(51.00,63.92)
\bezier{50}(51.00,63.92)(57.00,56.31)(63.00,55.66)
\bezier{50}(63.00,55.66)(69.00,54.64)(75.00,55.14)
\bezier{50}(75.00,55.14)(81.00,55.49)(87.00,55.76)
\bezier{50}(87.00,55.76)(93.00,55.97)(99.00,55.97)
\bezier{50}(99.00,55.97)(105.00,55.95)(111.00,56.01)
\bezier{50}(111.00,56.01)(117.00,56.08)(123.00,56.51)
\bezier{50}(123.00,56.51)(129.00,56.95)(135.00,57.95)
\bezier{50}(135.00,57.95)(141.00,58.96)(147.00,60.61)
\bezier{50}(147.00,60.61)(153.00,62.26)(159.00,64.55)
\bezier{50}(159.00,64.55)(165.00,66.85)(171.00,69.76)
\bezier{50}(171.00,69.76)(177.00,72.67)(183.00,76.15)
\bezier{50}(183.00,76.15)(189.00,79.62)(195.00,83.60)
\bezier{50}(195.00,83.60)(201.00,87.58)(207.00,92.00)
\bezier{50}(207.00,92.00)(213.00,96.43)(219.00,101.24)
\bezier{50}(219.00,101.24)(225.00,106.06)(231.00,111.22)
\bezier{50}(231.00,111.22)(237.00,116.38)(243.00,121.84)
\bezier{50}(243.00,121.84)(249.00,127.30)(255.00,133.03)
\bezier{50}(255.00,133.03)(261.00,138.75)(267.00,144.71)
\bezier{50}(267.00,144.71)(273.00,150.67)(279.00,156.83)
\bezier{50}(279.00,156.83)(285.00,163.00)(291.00,169.34)
\end{picture}
\end{center}
\caption[]{The ground state energy $2E^{(s)}_{0}$ as a function of
$\omega$ for $L=+2$, $2E_0({\rm exact})=1.18657366612601$,
$2\lambda=0.500008389662453$, $g=0.260754375208969$.}
\end{figure}
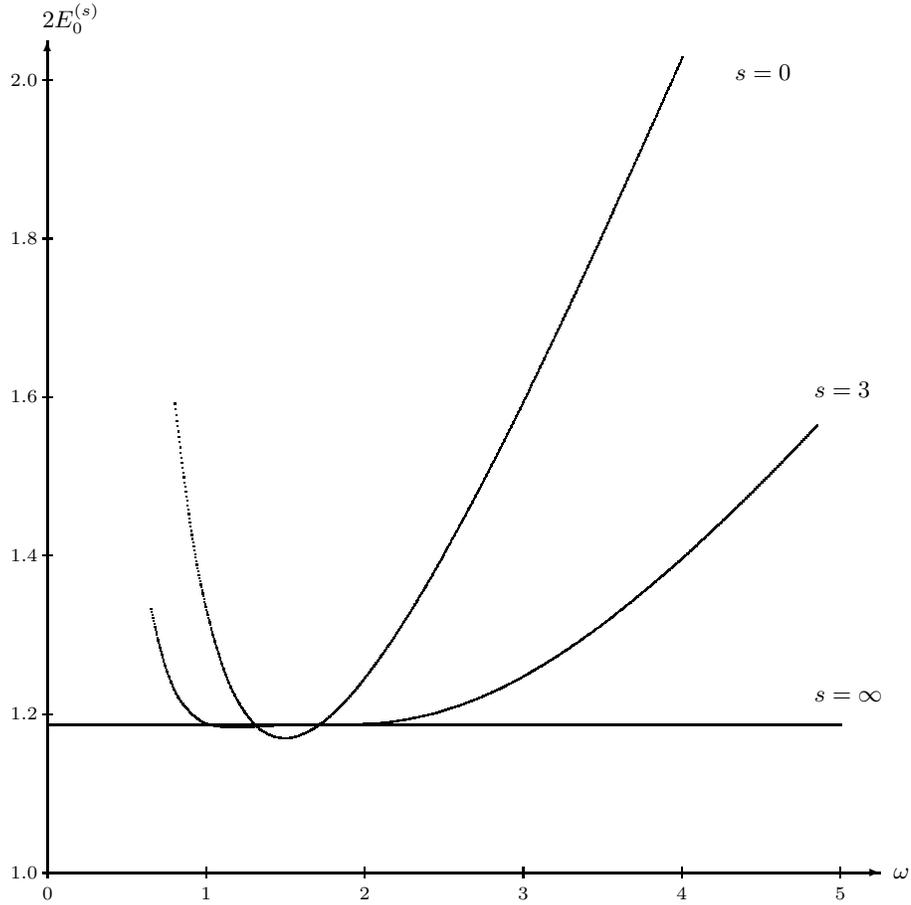
\end{document}